\documentclass[aps,preprint,showpacs,preprintnumbers,amsmath,amssymb]{revtex4}

\usepackage{dcolumn}
\usepackage{mathrsfs}
\usepackage{bm}
\usepackage{amsmath,amssymb,epsfig,float}

\begin{document}

\title{System-environment dynamics of X-type states in noninertial frames}

\author{Jieci Wang$^{1,2}$ and Jiliang Jing$^{1}$\footnote{Corresponding author, Email: jljing@hunnu.edu.cn}}
\affiliation{1) Department of Physics, and Key Laboratory of Low
Dimensional Quantum Structures and Quantum
Control of Ministry of Education,\\
Hunan Normal University, Changsha, Hunan 410081, China \\
2)  Department of Physics and Center of Physics, University of
Minho,
 \\Campus of Gualtar, Braga 4710-057, Portugal}

\vspace*{0.2cm}
\begin{abstract}
\vspace*{0.2cm} The  system-environment dynamics of noninertial systems
is investigated. It is shown that for the amplitude damping channel: (i) the biggest difference between the decoherence effect and the Unruh radiation on the dynamics of the entanglement is  the former only leads to entanglement transfer in the whole system, but the latter damages all types of entanglement; (ii) the system-environment entanglement increases and then declines, while the environment-environment entanglement always increases as the decay parameter $p$ increases; and (iii) the thermal fields generated by the Unruh effect can promote the sudden death of entanglement between the subsystems while postpone the sudden birth of entanglement between the environments. It is also found that  there is no system-environment and environment-environment  entanglements when the system coupled with the phase damping environment.
\end{abstract}

\vspace*{1.5cm}

\maketitle

\section{introduction}

It is well known that quantum entanglement is a key resource
for the implementation of many quantum information protocols \cite{Bouwmeester}, such as quantum communication, quantum cryptography, quantum  teleportation and computation. However, although many efforts have been made
to the study of the properties of entanglement, the good understanding of such a resource is only limited in inertial systems. Doubtlessly, the research of the entanglement behaviors in a relativistic setting will not only provide a more complete framework for the quantum information theory, but also play an important role in the understanding of the entropy and  information paradox \cite{Bombelli-Callen,
Hawking-Terashima} of  black holes. In addition, it is also closely related to
the implementation of quantum computation with observers in arbitrary relative motion \cite{David} and the study of the physical bounds of quantum information processing
tasks. As a result,  there are
an increasing number of articles discussing the entanglement
 in the relativistic setting, in particular on how the Unruh and
Hawking effect influence the degree of entanglement
\cite{Alsing-Mann, Schuller-Mann, Alsing-Milburn, moradi, Qiyuan, jieci1,jieci2,jieci3,
 Jason, Ralph, David, Andre, adesso, Eduardo, xiao,Khan} .

On the other hand, real quantum systems are necessarily subjected to their environments, and these reciprocal interactions often result in loss of quantum coherence and entanglement. Such a process is usually called quantum decoherence  \cite{Zurek,Breuer}, which has been widely investigated. It is often stated that the decoherence causes the system to become entangled with its environment,  and then the dynamics of the system is non-unitary.  It plays a fundamental role in the description of the quantum-to-classical transition \cite{Giulini,
Schlosshauer} and has been successfully applied in the cavity QED  \cite{Brune} and the ion trap experiments \cite{Myatt}.

In this paper we will discuss the system-environment dynamics for X-type state of the Dirac fields in a noninertial frame. As far as we known, either the entanglement behaviors  of the X-type states or the system-environment dynamics has not been investigated in noninertial frames yet.  The Dirac field, as shown in Refs. \cite{adesso2, E.M, Bruschi}, from an inertial perspective, can be described by a superposition of  Unruh  monochromatic modes $|0\rangle_{U}=\bigotimes_\omega|0_{\omega}\rangle_{U}
$ and $|1\rangle_{U}=\bigotimes_\omega|1_{\omega}\rangle_{U}$,  with
\begin{eqnarray}\label{Dirac-vacuum}
|0_{\omega}\rangle_{U} =
\cos r|0_{\omega}\rangle_{I}|0_{\omega}\rangle
_{II}+\sin r|1_{\omega}\rangle_{I}|1_{\omega}\rangle
_{II},
\end{eqnarray}
and
\begin{eqnarray}\label{Dirac-excited}
|1_{\omega}\rangle_{U}=|1_{\omega}
\rangle_{I}|0_{\omega}\rangle_{II},
\end{eqnarray}
where $\cos r=(e^{-2\pi\omega c/a}+1)^{-1/2}$, $a$ is the
acceleration of the observer, $\omega$
is frequency of the Dirac particle, and $c$ is the speed of light in
vacuum. We assume that two
observers, Alice and Rob, share an entangled
X-type initial state. Rob detects a single Unruh mode and Alice detects a
monochromatic Minkowski mode of the Dirac field.
Considering that an accelerated observer must remain in either region $I$
or $II$ due to these two regions are causally disconnected, i. e., an observer in region $I$ can't access to the field modes in the causally disconnected region
$II$,  we should trace over the inaccessible modes.

The outline of the paper is as follows. In Sec. II we recall some
concepts from the view of the quantum information theory, in
particular the theory of open quantum systems. In Sec. III we
investigate the system-environment dynamics of
X-type states in the noninertial frames. We summarize and discuss our
conclusions in the last section.

\section{Open system dynamics}

Let us start by briefly review the theory of open quantum systems (for details see Ref.  \cite{Brune}).
The total evolution of a system plus the environment  can
be described by $U_{SE}(\rho_S\otimes|0\rangle_E\langle 0|)U^\dag_{SE}$, where $U_{SE}$ is the evolution operator for the total state, and $|0\rangle_E$ represents the initial state of the
environment. By tracing over the degrees of freedom of the
environment, we can get the evolution  of
the system
\begin{eqnarray}\label{ff}
L(\rho_S)\begin{array}[t]{l}
=Tr_E[U_{SE}(\rho_S\otimes|0\rangle_E\!\langle 0|)U^\dag_{SE}]\;\\
=\sum_{\mu}  \!_E\langle \mu |U_{SE}|0\rangle_E\rho_S\,_E\!\langle 0| U^\dag_{SE}| \mu \rangle_E,
\end{array}
\label{TrKraus}
\end{eqnarray}
where $|\mu \rangle_E$ is the orthogonal basis for the environment,
and the operator $L$ describes the evolution of the system. Eq.
(\ref{ff}) can also be expressed as
\begin{eqnarray}
L(\rho_S)=\sum_\mu M_\mu \rho_S M_\mu^\dag, \label{EvolKraus1}
 \end{eqnarray}
 where
 \begin{eqnarray}\label{kraus}
 M_\mu\equiv\,
 _E\langle\mu | U_{SE}|0\rangle_E,
\end{eqnarray}
 are the Kraus operators~\cite{Kraus,Choi}. There are at most $d^2$
independent Kraus operators, where $d$ is the dimension of the system \cite{Salles, Leung03}.
Together with Eq. (\ref{kraus}), the dynamical evolution of the complete
system-environment state can be also given by the following map \cite{Salles} :%
\[
\mathbf{U}_{SE}|\xi_{l}\rangle_{S}\otimes|0\rangle_{E}=\sum_{k} M
_{k}|\xi_{l}\rangle_{S}\otimes|k\rangle_{E}\text{.}%
\]
\begin{widetext}with
\begin{align}
|\xi_{1}\rangle_{S}\otimes|0\rangle_{E} &  \rightarrow M _{0}|\xi
_{1}\rangle_{S}\otimes|0\rangle_{E}+\cdots+ M_{d^{2}-1}|\xi_{1}%
\rangle_{S}\otimes|d^{2}-1\rangle_{E}\nonumber\\
|\xi_{2}\rangle_{S}\otimes|0\rangle_{E} &  \rightarrow M_{0}|\xi
_{2}\rangle_{S}\otimes|0\rangle_{E}+\cdots+ M_{d^{2}-1}|\xi_{2}%
\rangle_{S}\otimes|d^{2}-1\rangle_{E}\nonumber\\
&  \vdots\nonumber\\
|\xi_{d}\rangle_{S}\otimes|0\rangle_{E} &  \rightarrow M_{0}|\xi
_{d}\rangle_{S}\otimes|0\rangle_{E}+\cdots+ M_{d^{2}-1}|\xi_{d}%
\rangle_{S}\otimes|d^{2}-1\rangle_{E}\text{,}\label{MapGen}%
\end{align}
\end{widetext}
where $\left\{  |\xi_{l}\rangle
_{S}\right\}  $ ( $l=1,\cdots,d$) is the complete basis for the system.

\section{System-environment dynamics of  entanglement}

We assume that Alice and Rob share a X-type initial state
\begin{equation}
\rho_{AR}=\frac{1}{4}\left(I_{AR}+ \sum_{i=0}^{3}c_{i}\sigma_{i}%
^{(A)}\otimes\sigma_{i}^{(R)}\right),
\label{initial}%
\end{equation}
where $I_{AR}$ is the identity operator in the Hilbert space of the
two qubits $A$ and $R$,  $\sigma_{i}^{(A)}$and $\sigma_{i}^{(R)}$ are the Pauli operators of the qubits $A$ and $R$, and $c_{i}$ ($0\leq \mid c_{i}\mid\leq1$) are real
numbers satisfying the unit trace and positivity conditions of the density
operator $\rho_{AR}$. Eq. (\ref{initial}) represents a class of states including the general initial state, the Werner initial state
($\left\vert c_{1}\right\vert =\left\vert c_{2}\right\vert =\left\vert
c_{3}\right\vert =c$), and the Bell basis ($\left\vert c_{1}\right\vert
=\left\vert c_{2}\right\vert =\left\vert c_{3}\right\vert =1$).
After the coincidence of Alice and Rob, Alice stays
stationary while Rob moves with uniform acceleration $a$. Using Eqs.
(\ref{Dirac-vacuum}) and (\ref{Dirac-excited}), we can rewrite Eq.
(\ref{initial}) in terms of Minkowski modes for Alice and Rindler
modes for Rob. Since Rob is causally disconnected from the region $II$, the only
information which is physically accessible to the observers is
encoded in the Minkowski modes $A$ described by Alice and the Rindler modes $\tilde{R}$
described by Rob. Taking the trace over the modes in region $II$, we
obtain
\begin{eqnarray}  \label{eq:state1}
\nonumber\rho_{A\tilde{R}}=\frac{1}{4}
\left[
  \begin{array}{cccc}
    (1+c_3)\cos^2 r & 0 & 0 &  c^-\cos r \\
    0 &(1+c_3)\sin^2 r+(1-c_3) &  c^+\cos r   & 0 \\
    0 &  c^+\cos r  & (1-c_3)\cos^2 r  & 0 \\
    c^-\cos r  & 0 & 0 & (1-c_3)+(1+c_3)\sin^2 r \\
  \end{array}
\right]
,
\end{eqnarray}
where $|mn\rangle=|m\rangle_{A}|n\rangle_{\tilde{R}}$, $c^+=c_1+c_2$, and $c^-=c_1-c_2$.

\subsection{Amplitude damping}

Now we  consider both Alice
and Rob's qubits under the amplitude damping  environment,
and the environment acts independently on Alice and Rob's states.
From Eq. (\ref{MapGen}) we find that the action of the amplitude damping channel over one qubit can be represented by the following phenomenological map \cite{Maziero}
\begin{subequations}
\begin{align}
|0\rangle_{i}|0\rangle_{E_i} &  \rightarrow|0\rangle_{i}|0\rangle_{E_i}%
\label{Map1}\\
|1\rangle_{i}|0\rangle_{E_i} &  \rightarrow\sqrt{q_i}|1\rangle_{i}|0\rangle
_{E_i}+\sqrt{p_i}|0\rangle_{i}|1\rangle_{E_i},\label{Map2}%
\end{align}
\end{subequations}
where  $q_i=1-p_i$, and $|0\rangle_{i}~(i=A,\tilde{R})$ are the ground and $|1\rangle_{i}$ are the excited qubit states of the $A\tilde{R}$ system. $|0\rangle_{E_i}$ and $|1\rangle_{E_i}$ describe the states of the environment with no excitation and one excitation of its
modes, respectively. We use $p_i$ ($0\leq \mid p_{i}\mid\leq1$) to describe
these probabilities as a parametrization of time. Here we only consider the simplest case in which all the
subsystems are embedded in the same environments, i.e.,
$p_{A}=p_{\tilde{R}}=p$  \cite{Salles}.

The total  initial  density operator of the whole system can be described as
\begin{equation}
\rho_{A\tilde{R}E_{A}E_{\tilde{R}}}=\rho_{A\tilde{R}}\otimes\rho_{E_AE_{\tilde{R}}},
\label{init}%
\end{equation}
where $\rho_{E_i}$ is the vacuum  state of the
environments.  Now by use of  Eqs. (\ref{Map1}), (\ref{Map2}) and (\ref{init}), we can compute the entanglement of the total density matrix $\rho_{A\tilde{R}E_{A}E_{\tilde{R}}}$ and discuss how it evolves. But here we are interesting in the entanglement dynamics of the bipartite subsystems (especially the system-environment dynamics), we only need to consider the corresponding bipartite reduced matrixes. The reduced-density matrix of the inertial subsystem $A$ and  the noninertial subsystem $\tilde{R}$, obtained by
taking the partial trace of $\rho_{A\tilde{R}E_{A}E_{\tilde{R}}}$ over the degrees of
freedom of the environment  $\rho_{A\tilde{R}}(a) =\operatorname*{Tr}%
\nolimits_{E_{A}E_{\tilde{R}}}[\rho_{A\tilde{R}E_{A}E_{\tilde{R}}}]$, is
given by \begin{widetext}
\begin{equation}
\rho_{A\tilde{R}}(a)=\frac{1}{4}%
\begin{bmatrix}
\alpha & 0 & 0 & qc^-\cos r \\
0 & q(\gamma+\beta p)  &qc^+\cos r & 0\\
0 & qc^+\cos r &q[\varepsilon +\beta p] & 0\\
qc^-\cos r  & 0 & 0 & \beta q^2
\end{bmatrix}
, \label{r1ad}%
\end{equation}
\end{widetext}
where $\alpha=\epsilon+p(2\varepsilon+\beta p) $, $\beta=(1+c_3)+\sin^2 r (1-c_3) $, $\gamma=(1-c_3)+\sin^2 r (1+c_3) $, $\varepsilon=(1-c_3)\cos^2 r$ and $\epsilon=(1+c_3)\cos^2 r$.

We are especially interested in the dynamical evolution of entanglement between  the
noninertial subsystem $\tilde{R}$ and its environment $E_{\tilde{R}}$.  The corresponding
reduced-density operator can be obtained by tracing
over the degrees of freedom of subsystem $A$ and  environment $E_{A}$ %
 \begin{widetext}
\begin{equation}
\rho_{\tilde{R}E_{\tilde{R}}}(a)=\frac{1}{4}%
\begin{bmatrix}
2\cos^{2} r  & 0 & 0 & 0 \\
0 & p(\beta+\gamma)  &  \sqrt{pq}(\beta+\gamma)& 0\\
0 &   \sqrt{pq}(\beta+\gamma)  & q(\beta+\gamma) &0\\
0  & 0 & 0& 0
\end{bmatrix}
, \label{r2ad}%
\end{equation}
\end{widetext}
and similarly, the state of the noninertial subsystem $\tilde{R}$ and the environment $E_A$ reads
 \begin{widetext}
\begin{equation}
\rho_{\tilde{R}E_A}(a)=\frac{1}{4}%
\begin{bmatrix}
\delta & 0 & 0 & \sqrt{pq}c^- \cos r \\
0 &q (\gamma+\beta q)   &\sqrt{pq}c^+ \cos r & 0\\
0 & \sqrt{pq}c^+ \cos r &p[\varepsilon +\beta p] & 0\\
\sqrt{pq}c^- \cos r  & 0 & 0 & \beta pq
\end{bmatrix}
, \label{r3ad}%
\end{equation}
\end{widetext}
where $\delta=\epsilon+q[\beta p+\varepsilon]+\gamma p$.

Again,  by tracing out the system degrees of freedom, we get the bipartite matrix of the partition $E_{A}E_{\tilde{R}}$
\begin{widetext}
\begin{equation}
\rho_{E_AE_{\tilde{R}}}=\frac{1}{4}%
\begin{bmatrix}
\chi& 0 & 0 &pc^-\cos r \\
0 & p(\gamma+\beta q)   &pc^+ \cos r & 0\\
0 & pc^+ \cos r &p[\varepsilon +\beta q] & 0\\
pc^- \cos r  & 0 & 0 & \beta p^2
\end{bmatrix}
, \label{r4ad}%
\end{equation}
\end{widetext}
where $\chi=\epsilon+q[\beta q+\varepsilon+\gamma]$.

\begin{figure}[ht]
\includegraphics[scale=0.65]{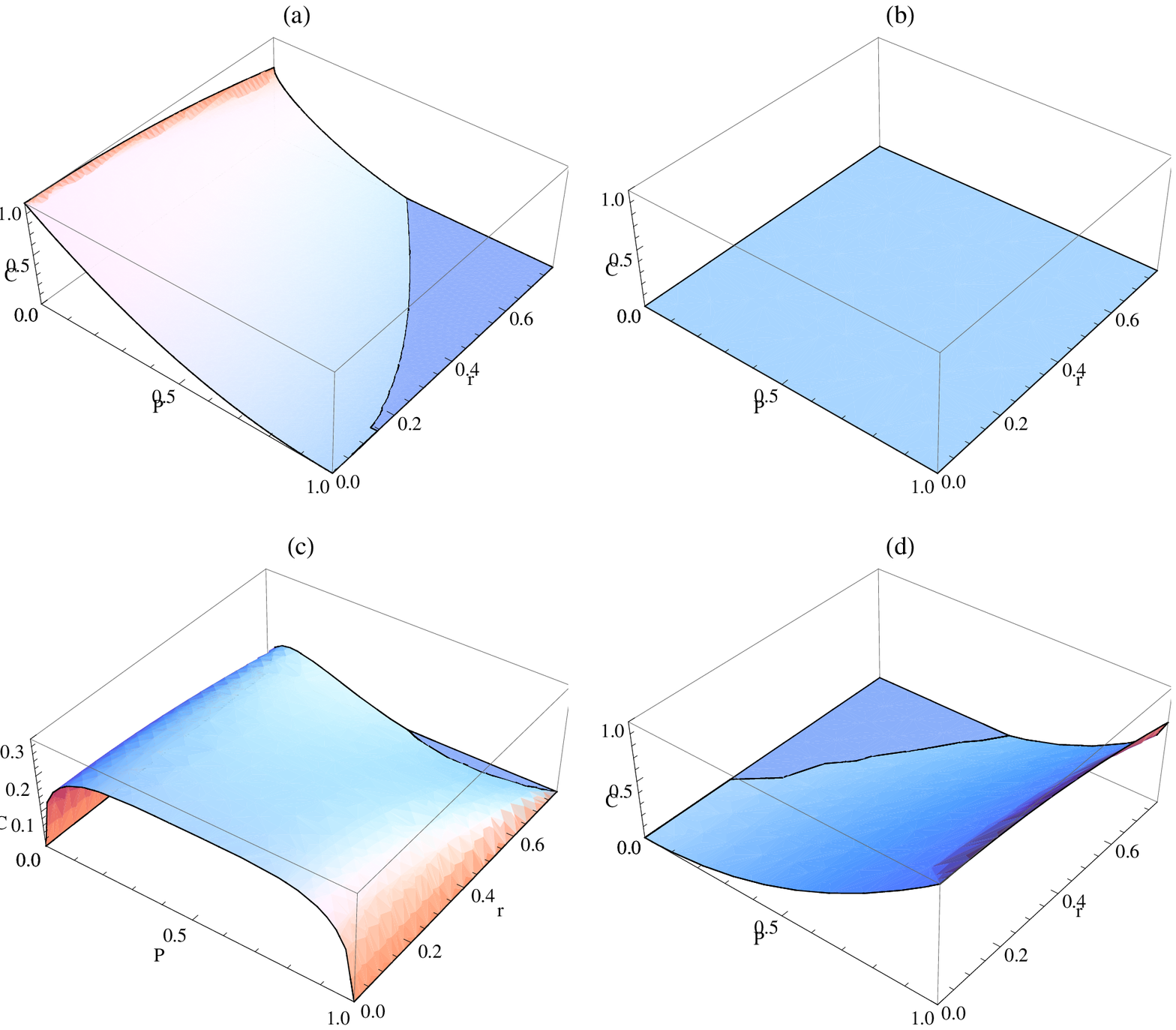}
\caption{\label{fig1}(Color online) Entanglement dynamics for the amplitude-damping
channel, considering the cases of Bell initial states ($\left\vert c_{1}\right\vert
=\left\vert c_{2}\right\vert =\left\vert c_{3}\right\vert =1$),
for bipartite states: (a) $A\tilde{R}$,  (b) $\tilde{R}E_{\tilde{R}}$, (c) $\tilde{R}E_{A}$, and
(d) $E_{A}E_{\tilde{R}}$ respectively.}
\end{figure}

\begin{figure}[ht]
\includegraphics[scale=0.75]{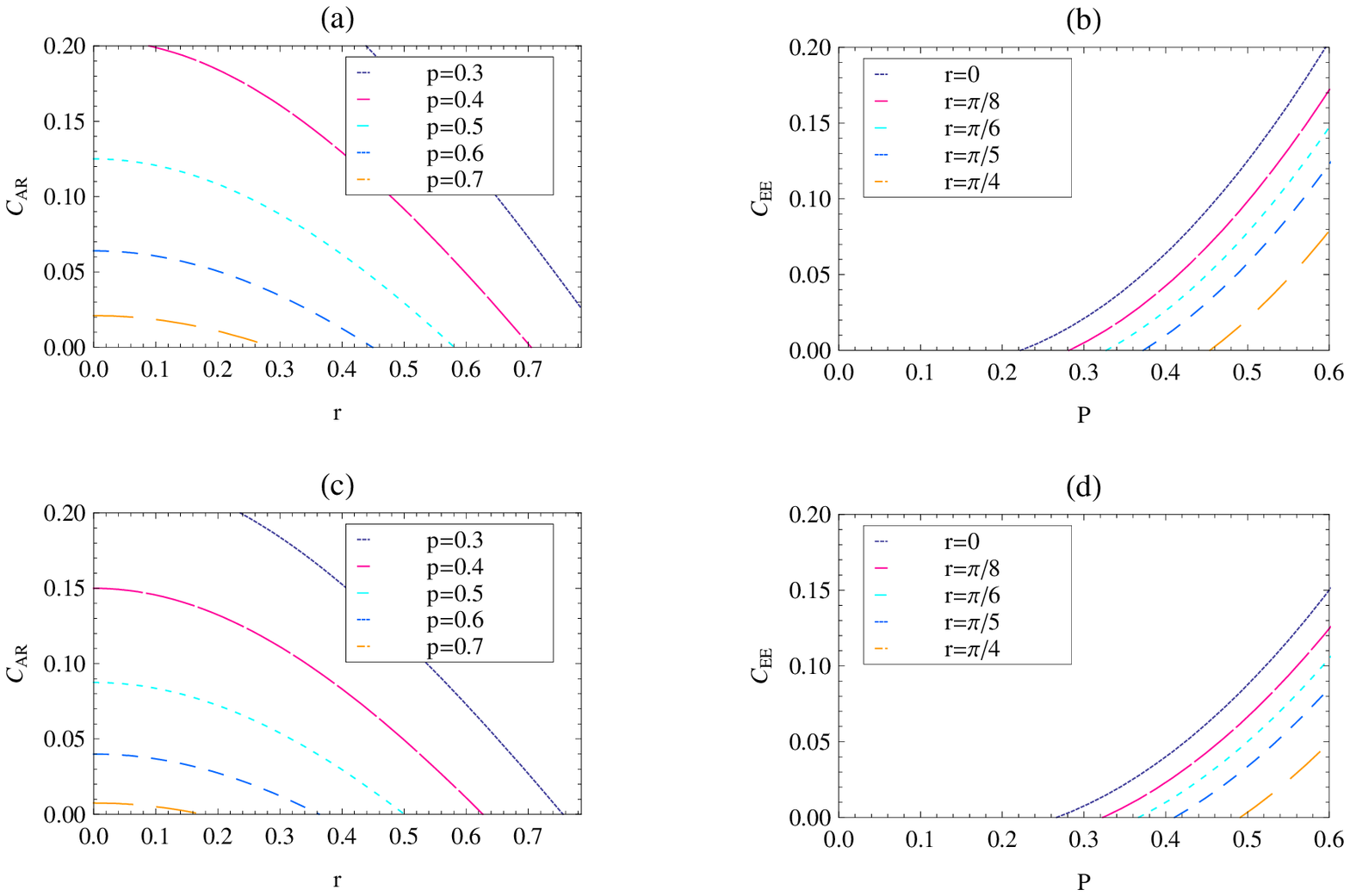}
\caption{\label{fig2}(Color online) Entanglement dynamics for the amplitude-damping
channel, considering the cases:
(a) bipartition $A\tilde{R}$ ($\left\vert c_{1}\right\vert
=\left\vert c_{2}\right\vert =\left\vert c_{3}\right\vert =0.8$),
 (b) bipartition $E_{A}E_{\tilde{R}}$ ($\left\vert c_{1}\right\vert
=\left\vert c_{2}\right\vert =\left\vert c_{3}\right\vert =0.8$), (c) bipartition $A\tilde{R}$  ($\left\vert c_{1}\right\vert
=0.7, \left\vert c_{2}\right\vert =0.9, \left\vert c_{3}\right\vert =0.4$), and (d) bipartition $E_{A}E_{\tilde{R}}$ ($\left\vert c_{1}\right\vert
=0.7, \left\vert c_{2}\right\vert =0.9, \left\vert c_{3}\right\vert =0.4$) respectively.}
\end{figure}

The entanglement of a
two-qubits mixed state $\rho$ in a noisy environments can be quantified
conveniently by the concurrence, which is defined as
\cite{Wootters,Coffman}
\begin{eqnarray}  \label{Concurrence}
C(\rho) =\max \left\{ 0,\lambda _{1}-\lambda
_{2}-\lambda _{3}-\lambda _{4}\right\}, \quad\lambda_i\ge
\lambda_{i+1}\ge 0,
\end{eqnarray}
where $\lambda_i$ are the square roots of the eigenvalues of the
matrix $\rho\tilde{\rho}$,
$\tilde{\rho}=(\sigma_y\otimes\sigma_y)\,
\rho^{*}\,(\sigma_y\otimes\sigma_y)$ is the ``spin-flip" matrix.
But fortunately,  due to the density matrixes from (\ref{r1ad}) to
(\ref{r4ad})  have $X$-type structures, here we have a simpler expression
for the concurrence \cite{Maziero}%
\begin{equation}
C(\rho)=2\max\left\{  0,\tilde{C}_{1}(\rho),\tilde{C}_{2}(\rho)\right\}  ,\label{Conc1}%
\end{equation}
with $\tilde{C}_{1}(\rho)=\sqrt{\rho_{14}\rho_{41}}-\sqrt{\rho_{22}\rho
_{33}}$ and $\tilde{C}_{2}(\rho)=\sqrt{\rho_{23}\rho_{32}}-\sqrt{\rho
_{11}\rho_{44}}$, where $\rho_{ij}$ are elements of the density matrix $\rho$. Then we can easily obtain analytical expressions of the
concurrence for the bipartite matrixes from (\ref{r1ad}) to (\ref{r4ad})
and plot them in the Figs. \ref{fig1}  and  \ref{fig2}. Note that the entanglement  of bipartite subsystems $AE_A(a)$ and $AE_{\tilde{R}}(a)$ are not presented in these figures. This is due to the fact that the environments $E_A$ and $E_{\tilde{R}}$ are symmetrical, thus the density matrix representing the bipartite subsystem $AE_A(a)$ is similar
to that of the bipartite subsystem $\tilde{R}E_{\tilde{R}}(a)$, and leading to  a similar
dynamic. In fact, we can prove that the concurrence $C_{AE_A(a)}=C_{\tilde{R}E_{\tilde{R}}(a)}=0$.  The same similarity exists between the
bipartite subsystems $AE_{\tilde{R}}(a)$ and $\tilde{R}E_{A}(a)$.

Fig.  \ref{fig1} shows the dynamics of the entanglement for all the partitions $A\tilde{R}$,  $\tilde{R}E_{\tilde{R}}$, $\tilde{R}E_{A}$, and $E_{A}E_{\tilde{R}}$, when $\left\vert c_{1}\right\vert=\left\vert c_{2}\right\vert =\left\vert c_{3}\right\vert =1$. It is shown
that the entanglement of the subsystem  $A\tilde{R}$ suffers
sudden death (SD) at some certain acceleration parameter $r$  and decay parameter $p$. We also note that the entanglement between the noninertial subsystem $\tilde{R}$ and its environment $E_{\tilde{R}}$  always equals to zero for any $r$ and $p$. However, the interaction between the system and environment generates  system-environment entanglement between the noninertial subsystem $\tilde{R}$ and the environment $E_A$. As the decay parameter $p$ increases, the system-environment entanglement of  $\tilde{R}E_A$  increases firstly and then decreases quickly. However, as the acceleration increases, such entanglement always decreases and appears SD at some larger accelerations.  At the same time, it is interesting to note that this interaction also generates environment-environment entanglement between the environments $E_{A}$ and $E_{\tilde{R}}$, and such entanglement
exhibits entanglement sudden birth (SB) \cite{Solano} at some certain $r$ and $p$. It is worthy to notice  that when $p=0$, the entanglement  of the system (Fig. 1a) is 1 while the system-environment entanglement (Fig. 1c) and environment-environment entanglement (Fig. 1d) are zero. However, when  $p=1$, there is only environment-environment entanglement. That is to say, at first the entanglement of the system was transferred to system-environment  and environment-environment entanglement, but finally all these lost entanglement were transferred to the environment degrees of freedom. Thus, we can see that  the system-environment entanglement increases, reaches a maximum, and then declines. Then we conclude that the most different between the decoherence and Unruh effect on the dynamics of the entanglement in noninertial frames is that the former only leads to  entanglement transfer in the whole system, while the latter damages not only entanglement of the system, but also system-environment and environment-environment entanglement. It is also shown that the SD of entanglement in Figs. 1a and 1c  occurs  almost as the same time as the SB of entanglement in Fig. 1d for very large $r$. In fact, it is very easy to show that SB might be occurring before, simultaneously with or even after SD, depending on different initial states. For example, one can plot a similar figure as Fig. 1 for the case of a Werner initial state with $\left\vert c_{1}\right\vert=\left\vert c_{2}\right\vert =\left\vert c_{3}\right\vert =0.7$ and find that  SB occurs much later than SD.

Fig. \ref{fig2} shows how the acceleration and decay parameter affect the SD and SB of the entanglement for Werner ($\left\vert c_{1}\right\vert=\left\vert c_{2}\right\vert =\left\vert c_{3}\right\vert =0.8$) and general ($\left\vert c_{1}\right\vert=0.7,  \left\vert c_{2}\right\vert =0.9,  \left\vert c_{3}\right\vert =0.4$)  initial states. We note that for both of these two cases: (i) the monotonous decrease of entanglement of the system $A\tilde{R}$ as the acceleration increase can attribute to the thermal fields generated by the Unruh effect;  (ii)  a larger $p$ leads to an earlier SD as the acceleration increases; and (iii) the entanglement between the environments $E_{A}$ and $E_{\tilde{R}}$ always increases  as time $p$ increases but decreases as the acceleration increases. However, it is worthy to note that a larger acceleration results in an earlier SD of entanglement between the subsystems $A$ and $\tilde{R}$ but a later SB of entanglement between the environments $E_{A}$ and $E_{\tilde{R}}$.  The thermal fields generated by the Unruh effect can promote SD but postpone SB of entanglement in noninertial frames. We also note that in the case of $p=0.3$ (and naturally when $p<0.3$), for a Werner initial state the entanglement didn't tends to zero even the acceleration approaches to infinity, which is quite different from the general initial state case. It seems that the form of initial state also plays an important role in the system-environment dynamics of the entanglement in noninertial frames.

\subsection{Phase damping channel}

In this subsection we discuss the dynamics of system-environment entanglement under the  phase-damping channel, which describes the loss of quantum coherence without losing
energy \cite{NieChu}. The map  of this channel on a one-qubit
system is given by%
\begin{subequations}
\begin{align}
|0\rangle_{i}|0\rangle_{E_i} &  \rightarrow|0\rangle_{i}|0\rangle_{E_i}%
\label{Mapl1}\\
|1\rangle_{i}|0\rangle_{E_i} &  \rightarrow\sqrt{q_i}|1\rangle_{i}|0\rangle
_{E_i}+\sqrt{p_i}|1\rangle_{i}|1\rangle_{E_i},\label{Mapl2}%
\end{align}
\end{subequations}
where $|0\rangle_{i}~(i=A,\tilde{R})$ .

The reduced-density operator for the partition $A\tilde{R}$, obtained by taking the
partial trace of $\rho_{A\tilde{R}E_{A}E_{\tilde{R}}} (p)$ over the degrees of
freedom of the environment, is
given by
\begin{widetext}
\begin{equation}
\rho_{A\tilde{R}} (p)=\frac{1}{4}%
\begin{bmatrix}
\epsilon & 0 & 0 & qc^- \cos r \\
0 &\gamma  &qc^+ \cos r & 0\\
0 & qc^+ \cos r &\varepsilon & 0\\
qc^- \cos r  & 0 & 0 & \beta
\end{bmatrix}
, \label{a1ad}%
\end{equation}
\end{widetext}

For the bipartite subsystems  $\tilde{R}E_{\tilde{R}}$ and $\tilde{R}E_{A}$,
the reduced-density operators are found to be%
\begin{widetext}
\begin{equation}
\rho_{\tilde{R}E_{\tilde{R}}} (p)=\frac{1}{4}%
\begin{bmatrix}
2\cos^2 r & 0 & 0 & 0 \\
0 & 0  &  0& 0\\
0 &  0  &q (\beta+\gamma) & \sqrt{pq}(\beta+\gamma)\\
0  & 0 & \sqrt{pq}(\beta+\gamma) &p (\beta+\gamma)
\end{bmatrix}
. \label{a2ad}%
\end{equation}
\end{widetext}

and

 \begin{widetext}
\begin{equation}
\rho_{\tilde{R}E_A} (p)=\frac{1}{4}%
\begin{bmatrix}
\varepsilon q+\epsilon & 0 & \sqrt{pq}\varepsilon & 0 \\
0 & \gamma+\beta q  &0 &  \beta \sqrt{pq}\\
\sqrt{pq}\varepsilon &0 &\varepsilon p & 0\\
0  & \beta \sqrt{pq}& 0 & \beta p^2
\end{bmatrix}
. \label{a3ad}%
\end{equation}
\end{widetext}

\begin{figure}[ht]
\includegraphics[scale=0.65]{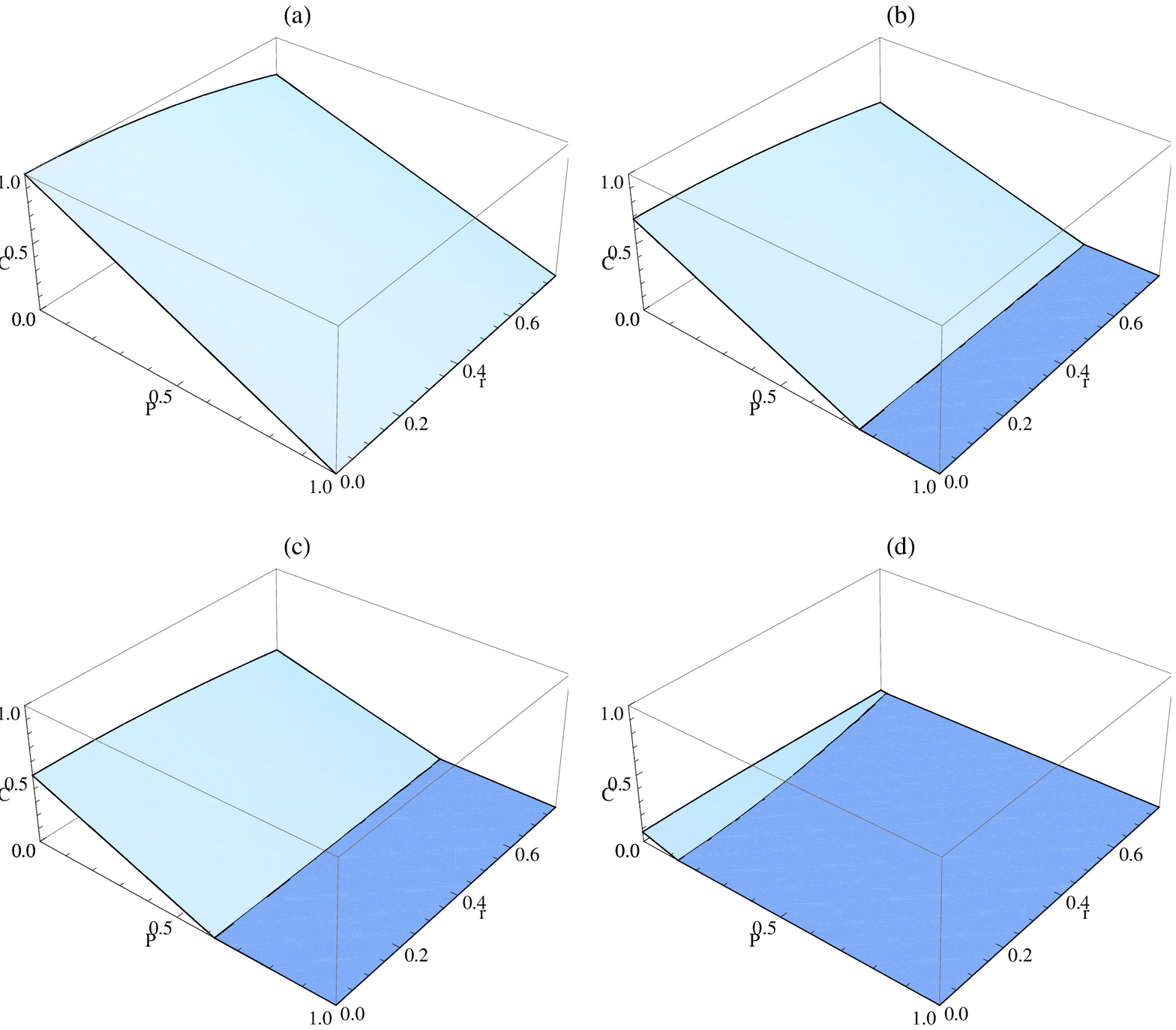}
\caption{\label{fig3}(Color online) Entanglement dynamics of bipartition $A\tilde{R}$ for the  phase damping channel, considering the cases: (a) Bell initial states, (b) Werner initial states $\left\vert c_{1}\right\vert
=\left\vert c_{2}\right\vert =\left\vert c_{3}\right\vert =0.9$, (c) Werner initial states $\left\vert c_{1}\right\vert=\left\vert c_{2}\right\vert =\left\vert c_{3}\right\vert =0.8$,
and (d) general initial states $\left\vert c_{1}\right\vert=0.6, \left\vert c_{2}\right\vert =0.5,
\left\vert c_{3}\right\vert =0.3$ respectively.}
\end{figure}

Similarly, for the partition $E_{A}E_{\tilde{R}} (p)$,  by tracing out the system
degrees of freedom, we obtain
\begin{widetext}
\begin{equation}
\rho_{E_AE_{\tilde{R}}} (p)=\frac{1}{4}%
\begin{bmatrix}
\varpi & \sqrt{pq}(\gamma+\beta q) & \sqrt{pq}(\varepsilon+\beta q)  &\beta pq \\
\sqrt{pq}(\gamma+\beta q) & (\gamma+\beta q) p  &\beta pq & \beta p\sqrt{pq}\\
\sqrt{pq}(\varepsilon+\beta q)   & \beta pq &p[\varepsilon +\beta q] & \beta p\sqrt{pq}\\
\beta pq  & \beta p\sqrt{pq} & \beta p\sqrt{pq} & \beta p^2
\end{bmatrix}
, \label{a4ad}%
\end{equation}
\end{widetext}
where $\varpi=\epsilon+q[\beta q+\varepsilon+\gamma]$. We can see that only the density matrix Eq. (\ref{a1ad})   has an $X$-type structure. By using of the Peres separability criterion \cite{Peres1}, we find that there is no entanglement in bipartite states from Eqs.    (\ref{a2ad}) to  (\ref{a4ad}).  In other words, the interaction between system and environment didn't generate bipartite system-environment and environment-environment entanglement  in the phase damping case.

Fig. \ref{fig3} shows how the acceleration and decay parameter change the entanglement
of the system $A\tilde{R}$ for different initial states. We find again that the entanglement of the system, as well as the system-environment and environment-environment entanglement decrease as $r$ increases for fixed $p$, which is as same as the amplitude damping case. We can see that (i) for a Bell state ($\left\vert c_{1}\right\vert=\left\vert c_{2}\right\vert =\left\vert c_{3}\right\vert =1$), there is no SD of entanglement, which is quite different from the amplitude-damping case;  (ii) for Werner states ($\left\vert c_{1}\right\vert=\left\vert c_{2}\right\vert =\left\vert c_{3}\right\vert =0.9$) and ($\left\vert c_{1}\right\vert=\left\vert c_{2}\right\vert =\left\vert c_{3}\right\vert =0.8$) the SD always appears  as $r$ and $p$ increase;
and (iii) for general initial states ($\left\vert c_{1}\right\vert=0.6,  \left\vert c_{2}\right\vert =0.5,  \left\vert c_{3}\right\vert =0.3$), the SD of entanglement appears very early.  Now we can safely come to the conclusion that the form of initial state do plays an important role in the system-environment dynamics of entanglement in noninertial frames.

\section{summary}

We investigated the system-environment dynamics in a noninertial  frame when both the noninertial and inertial subsystems coupled with environments. It is shown that for the amplitude damping channel, only the entanglement between subsystem $\tilde{R}$ and its environment $E_{\tilde{R}}$ equals to zero. However,  there is no entanglement  in bipartite states $\tilde{R}E_{\tilde{R}}$, $\tilde{R}E_{A}$, and $E_{A}E_{\tilde{R}}$ when the system coupled with the phase damping environment.  It is found  that the  biggest difference between the decoherence and Unruh effect on the dynamics of the entanglement in noninertial frames is the former only leads to  entanglement transfer in the whole system, while the latter damages not only entanglement of the system, but also system-environment and environment-environment entanglement.  In the amplitude damping case,  the entanglement of the system  $A\tilde{R}$ suffers SD at some certain $p$ and $r$ for any initial states. However, when the system is under the phase damping channel, for a Bell initial state the entanglement tends to zero only when the acceleration approaches to infinity or $p=1$, which is quite different from the amplitude-damping case.
At the same time, it is found that for the amplitude damping, the interaction between system and environment generates bipartite system-environment entanglement between the noninertial subsystem $\tilde{R}$ and the environment $E_A$. As the decay parameter $p$ increases, the system-environment entanglement  increases firstly and then decreases quickly. However, as the acceleration increases, the system-environment entanglement always decreases and appears a SD at some larger accelerations.   It is interesting to note that when the system coupled with the amplitude damping environment, such an interaction also generates entanglement between the environments $E_{A}$ and $E_{\tilde{R}}$, and such entanglement exhibits a SB.   It is worthy to notice that a larger acceleration results in an earlier SD of entanglement between the subsystems $A$ and $\tilde{R}$ but a later SB of entanglement between the environments $E_{A}$ and $E_{\tilde{R}}$.  The thermal fields generated by the Unruh effect can promote SD but postpone SB of entanglement in  noninertial frames.  We  also find that the form of initial state  plays an important role in the system-environment dynamics of entanglement in noninertial frames.

\begin{acknowledgments}

This work was supported by the National Natural Science Foundation of China under Grant No. 11175065, 10935013;  PCSIRT, No. IRT0964; the Hunan Provincial Natural Science Foundation of China under Grant No 11JJ7001; he Hunan Provincial Innovation Foundation
For Postgraduate under Grant No. CX2010B216; and Construct Program of the National  Key Discipline. We thank the Kavli Institute for Theoretical Physics China for hospitality in the revised stages of this work. 	
\end{acknowledgments}

\end{document}